\documentclass[a4paper,11pt]{article}
\usepackage{pos}
\usepackage{float}

\usepackage{hyperref}
\usepackage{multirow}
\usepackage{lineno}

\usepackage{booktabs} 

\title{Testing high energy neutrino emission from the Fermi Gamma-ray Space Telescope Large Area Telescope (4LAC) sources}
 \ShortTitle{Testing high energy neutrino emission}

\author*[a]{Antonio Galv\'an}
\author[a]{Nissim Fraija}
\author[a]{Edilberto Aguilar-Ruiz}
\author[b]{Jagdish C. Joshi}
\author[a]{Jose Antonio de Diego Onsurbe}
\author[d]{Antonio Marinelli}

\affiliation[a]{Instituto de Astronom\'ia, Universidad Nacional Aut\'onoma de M\'exico, Ciudad de Mexico, Mexico}
\affiliation[b]{School of Astronomy and Space Science, Nanjing University, Nanjing210093, China}
\affiliation[c]{Key Laboratory of Modern Astronomy and Astrophysics (Nanjing Univer-sity), Ministry of Education, China}

\affiliation[d]{INFN - Sezione di Napoli, Complesso Univ. Monte S. Angelo, I-80126 Napoli, Italy}

\emailAdd{agalvan@astro.unam.mx}
\emailAdd{nifraija@astro.unam.mx}

\abstract{The detection of the high-energy neutrino IC-170922A in spatial (within the error region) and temporal flare activity correlation with the blazar TXS 0506+056 allowed these objects to be considered as progenitor sources of neutrinos. Besides this, no more detection of this kind was reported. Some other neutrinos detected by IceCube show a spatial correlation (within the error region) from other Fermi-LAT detected sources. However, these objects did not show a flare activity like TXS 0506+056. Assuming a lepto-hadronic scenario through p$\gamma$ interactions, this work describes the SED in some objects from the fourth catalog of active galactic nuclei (AGNs) detected by the Fermi Gamma-ray Space Telescope Large Area Telescope (4LAC) sources, which are in spatial correlation with neutrinos detected by IceCube. Additionally, we estimate the corresponding neutrino flux counterpart from these sources.
}

\FullConference{37$^{\rm{th}}$ International Cosmic Ray Conference (ICRC 2021)\\
		July 12th -- 23rd, 2021\\
		Online -- Berlin, Germany}

\begin{document}
\maketitle
\section{Introduction}
	\paragraph*{}
	Cosmic rays (high-energy protons or heavy nuclei) frequently hit the earth with energies greater than $>10^{18}$ eV. More than a century since the discovery, places, and processes of acceleration remain in uncertainty. This is due to the electric charge makes that any magnetic field deflects its trajectories to  Earth \cite{COUTU20121355}. Fortunately, theoretical models suggest that in their places of acceleration, cosmic rays interact with radiation fields of low energy and with matter to produce high-energy photons and neutrinos \cite{2019ARNPS..69..477M}. Besides the fact that cosmic rays cannot tell us their production places, gamma rays and neutrinos can. At very high energies (hundred of GeVs), photons are attenuated by the effect of Extragalactic Background Light (EBL) \citep[e.g.][]{2017A&A...603A..34F}. Therefore, a neutrino is an ideal messenger.


\section{Observations}\label{}

We use the muon neutrinos detected by IceCube up to date. For the electromagnetic counterpart, we use the 4LAC-DR2 catalog reported by Fermi-LAT. 

\subsection{The neutrino sample} \label{ch:nu}

The IceCube telescope is located near the Amundsen-Scott South Pole Station. This telescope is buried below the surface, reaching a depth of 2,500 meters. The IceCube Collaboration has made public the information about neutrinos detected principally in two catalogs: The High Energy Starting Event (HESE)  \citep{2013Sci...342E...1I, 2014PhRvL.113j1101A, 2015ICRC...34.1081K}, in which are reported 86 high-energy neutrinos, in addition to 36 neutrino events in the Extremely High Energy (EHE) catalog  \cite{2016ApJ...833....3A, 2017ICRC...35.1005H}. Additionally, with the goal of having a rapid response to electromagnetic transients possible associated to neutrinos, a real time alert system was implemented. This system distributes the real time events HESE\footnote{Those alerts are public and can be accessed in \url{https://gcn.gsfc.nasa.gov/amon_hese_events.html}} and EHE\footnote{Those alerts are public and can be accessed in \url{https://gcn.gsfc.nasa.gov/amon_ehe_events.html}} via AMON consortium \cite{2013APh....45...56S}. 

\subsection{The 4LAC sample} \label{ch:lat}
	
The Fermi satellite was launched in June 2008 and since then scan the sky constantly. The satellite is equipped with two instruments, the Gamma-Ray Burst Monitor (GMB)  \cite[]{2009ApJ...702..791M} and the Large Area Telescope (LAT) \cite[]{2009ApJ...697.1071A} instruments. Recently, the LAT team released the 4FGL catalog \citep[]{2020ApJS..247...33A, 2020arXiv200511208B}, which contains the sources in the sky that emits photons in the energy range from 50 MeV up to 1 TeV covering the first 10 years of operation.\\

\noindent We limit ourselves to test the sub set of Active Galactic Nuclei catalog (4LAC) \cite{2020ApJ...892..105A, 2020arXiv201008406L}, which contains AGN detected by Fermi-LAT with a galactic latitude ($|b| > 10^{\circ}$).  It compromises 3125 sources: 694 FSRQ, 1125 BL Lacs, 1240 BCU and 66 Non blazar AGN. Formerly, a sub sample that does not belong to  4LAC  known as Low Latitude sample ($|b| < 10^{\circ}$) is considered. This sub sample is constituted by 36 FSRQ, 65 BL Lacs, 260 BCU and 6 non blazar AGNs. In particular,  we choose the blazar population of this sample to be associated with high-energy neutrinos.

\section{Theoretical Model}\label{sec:HAWCResults}

The Spectral Energy Distribution of the blazar was dominate by non-thermal emission characterized by a double bump. The first one located at low energies (from Radio up to Soft X-rays) is interpreted  by synchrotron radiation produced by a population of relativistic electrons ($N_{e}$) embedded into a magnetic field ($B$) \citep[e.g.][]{peterson1997introduction,2007Ap&SS.309...95B}. Meanwhile, at higher energies, the mechanism remains in discussion. In a leptonic scenario, the same population of relativistic electrons interacts with the synchrotron photons. Relativistic electrons scatter these photons up to higher energies via the Inverse Compton process (SSC) \citep[e.g.][]{2017ApJS..232....7F, 2020arXiv201101847A}. On the other hand, hadronic processes can take place with the presence of cosmic rays in the plasma that interacts with radiation fields of low energies (p$\gamma$) \citep[e.g.][]{2017APh....89...14F,2016APh....80..115P}. In hadronic processes, neutral pions ($\pi^{0}$) and charged pions ($\pi^{\pm}$) are expected \cite{romero2013introduction}. In the case of the $\pi^{0}$, the channel $\pi^{0} \rightarrow \gamma \gamma$ takes place, while for charged pions the channels $\pi^{+} \rightarrow \mu^{+} + \nu_{\mu}$, and therefore $\mu^{+} \rightarrow e^{+} + \nu_{e} + \bar{\nu}_{\mu}$. From this, it is expected a neutrino rate of ($\nu_{\mu} : \nu_{e} : \nu_{\tau}$) = (2:1:0) in the source. In average the energy of the $\gamma$-rays is $E_{\gamma} \sim 1/2E_{\pi^{0}}$ and the neutrinos carry a fraction of energy $E_{\nu} \sim 1/4E_{\pi}$ from the parent pions, respectively \citep{2018PrPNP.102...73A}.
	
\subsection{Leptonic Scenario}
	



We adopt the theoretical framework derived by  \cite{2008ApJ...686..181F}.  The synchrotron spectra produced by an electron population with a profile $N_{e}^{\prime}(\gamma^{\prime})$ is described by:

\begin{equation}
    f_{syn}(\epsilon) =  \frac{\sqrt{3}\delta_{D}^{4}\epsilon^{\prime}e^{3}B}{4\pi h d_{L}^{2}} \int_{1}^{\infty} N_{e}^{\prime}(\gamma^{\prime})R(x) d\gamma^{\prime},
\label{eq:Synch}  
\end{equation}

\noindent where $\epsilon = h\nu/(m_{e}c^{2})$ is the frequency normalized at the electron mass, $\delta_{D}$ is the Doppler factor, $e$ the fundamental charge, $h$ the Planck's constant, $d_{L}$ the luminosity distance from the source and $R(x)$ is defined in \cite{2008ApJ...686..181F}. In the high-energy regime, the SSC spectrum is described by:

\begin{equation}
  f_{ssc}(\epsilon_{s}) =  \frac{9}{16}\frac{(1+z)^{2}\sigma_{T}\epsilon_{s}^{\prime 2}}{\pi \delta_{D}^{2} c^{2} t_{v, min}^{2}} \int_{0}^{\infty} \frac{f_{syn}(\epsilon)}{\epsilon^{\prime}} d\epsilon^{\prime} \int_{\gamma_{min}^{\prime}}^{\gamma_{max}^{\prime}} \frac{N_{e}^{\prime}(\gamma^{\prime}) }{\gamma^{\prime 2}} F_{c}(q, \Gamma_{e}) d\gamma^{\prime}\,, 
  \label{eq:SSC}
\end{equation}

\noindent where $z$ is the redshift, $\sigma_{T}$ is the Thompson cross section, $t_{var, min}$ is the variability timescale and $F_{C}(q, \Gamma_{e})$ is the Compton cross section \citep{1968PhRv..167.1159J}.

\subsection{Hadronic Scenario}





We adopt the hadronic scenario introduced in \cite{2008PhRvD..78c4013K}. A  proton population  described by their characteristic function $f_{p}(E_{p})$ and the target photon population $f_{ph}(\epsilon)$ is used. We adopt a proton population described by a power law with exponential cut-off and and the target photons described by a blackbody profile with energy $kT$. Therefore, the $\gamma$-ray spectra (and secondary leptons $l \in \{e,e^{+},\nu_{\mu} \bar{\nu}_{\mu}, \nu_{e} \bar{\nu}_{e} \}$) are described by the equation:



\begin{equation}
    \frac{dN}{dE} = \int_{\eta_{0}}^{\infty} H(\eta, E) d\eta\,,
    \label{eq:Hadronic}
\end{equation}

\noindent where $H(\eta, E)$ is defined as:

\begin{equation}
    H(\eta, E) = \frac{m_{p}^{2}c^{4}}{4} \int_{E}^{\infty}  \frac{f_{p}(E_{p})}{E_{p}^{2}} f_{ph}\left(\frac{\eta m_{p}^{2}c^{4}}{4E_{p}} \right) \Phi_{\gamma} \left(\eta, \frac{E}{E_{p}} \right) dE_{p}\,,
\end{equation}

\noindent the function $\Phi_{\gamma}$ (also $\Phi_{l}$) is a parametrization of the cross section of the hadronic processes given by \cite{2008PhRvD..78c4013K} over $\eta$. It is defined over the region where this kinematics interactions are allowed.
	
	
	
	


\section{Correlations founded} 

We apply a spatial correlation based on the angular distance between the high-energy neutrino (track events) detected by IceCube (section \ref{ch:nu}) and the Fermi-LAT blazars reported in 4LAC (section \ref{ch:lat}) considering that a gamma-ray source can be associated with a neutrino spatially if the position of this source relies in the neutrino error region at 90$\%$. The correlations that we found are reported in the Tables \ref{tab:CorrHESEEHE}, \ref{tab:CorrAMON} and \ref{tab:CorrGoldBronze}. In each table, we report the spatial correlations founded.  Table \ref{tab:CorrHESEEHE} shows the coincidences of the neutrinos reported in the HESE and the EHE catalogs, meanwhile in Table \ref{tab:CorrAMON} is shown the HESE and EHE alerts reported via AMON.  Finally, Table \ref{tab:CorrGoldBronze} is reported the association founded with the Gold and Bronze alerts.

\section{Implementation}



We pretend to adjust this model through the $\chi^{2}$ technique, by adopting the strategy proposed in \cite{2008ApJ...686..181F}. In an iterative process on the magnetic field (B) and the Doppler factor ($\delta_{D}$), we find the best values for the Synchrotron and SSC spectra assuming a leptonic scenario, using equations (\ref{eq:Synch}) and (\ref{eq:SSC}) respectively. Is is worth noting that if an extra component at gamma-rays regime is needed, a hadronic component will be considered through equation (\ref{eq:Hadronic}).


\section{Conclusion}\label{sec:ConADis}

We found spatial correlations between high-energy neutrinos (track events) reported by IceCube and blazars that emit gamma-ray photons at high energies. Quasi simultaneous SEDs are being derived around the neutrino arrival times  in order to describe them with a same radiative model. Now, we are working in Bayesian blocks \cite{2013ApJ...764..167S}, which are being applied to the GeV-TeV light curves in order to find flare episodes around the neutrino arrival times.




\begin{table}[]
\centering
\scalebox{0.65}{
\begin{tabular}{@{}ccccccc@{}}
\toprule
\multicolumn{7}{r}{\multirow{2}{*}{HESE}} \\
\multicolumn{7}{r}{} \\ \midrule
\multicolumn{1}{|c|}{ID} & \multicolumn{1}{c|}{Energy (TeV)} & \multicolumn{1}{c|}{MJD} & \multicolumn{1}{c|}{RA (deg)} & \multicolumn{1}{c|}{Dec (deg)} & \multicolumn{1}{c|}{Ang Err (deg)} & \multicolumn{1}{c|}{} \\ \midrule
\multicolumn{1}{|c|}{5} & \multicolumn{1}{c|}{71.4} & \multicolumn{1}{c|}{55512.55} & \multicolumn{1}{c|}{110.6} & \multicolumn{1}{c|}{-0.4} & \multicolumn{1}{c|}{1.2} & \multicolumn{1}{c|}{} \\ \midrule
\multicolumn{2}{|c|}{4FLG ASSOC} & \multicolumn{1}{c|}{ASSOC} & \multicolumn{1}{c|}{RA (deg)} & \multicolumn{1}{c|}{Dec (deg)} & \multicolumn{1}{c|}{Ang Sep (deg)} & \multicolumn{1}{c|}{Type} \\ \midrule
\multicolumn{2}{|c|}{*4FGL J0725.8-0054} & \multicolumn{1}{c|}{PKS 0723-008} & \multicolumn{1}{c|}{111.46} & \multicolumn{1}{c|}{-0.91} & \multicolumn{1}{c|}{1.01} & \multicolumn{1}{c|}{bcu} \\\midrule
\multicolumn{1}{|c|}{ID} & \multicolumn{1}{c|}{Energy (TeV)} & \multicolumn{1}{c|}{MJD} & \multicolumn{1}{c|}{RA (deg)} & \multicolumn{1}{c|}{Dec (deg)} & \multicolumn{1}{c|}{Ang Err (deg)} & \multicolumn{1}{c|}{} \\ \midrule
\multicolumn{1}{|c|}{23} & \multicolumn{1}{c|}{82.2} & \multicolumn{1}{c|}{55949.56} & \multicolumn{1}{c|}{208.7} & \multicolumn{1}{c|}{-13.2} & \multicolumn{1}{c|}{1.9} & \multicolumn{1}{c|}{} \\ \midrule
\multicolumn{2}{|c|}{4FLG ASSOC} & \multicolumn{1}{c|}{ASSOC} & \multicolumn{1}{c|}{RA (deg)} & \multicolumn{1}{c|}{Dec (deg)} & \multicolumn{1}{c|}{Ang Sep (deg)} & \multicolumn{1}{c|}{Type} \\ \midrule
\multicolumn{2}{|c|}{4FGL J1359.1-1152} & \multicolumn{1}{c|}{2MASS   J13592131-1150440} & \multicolumn{1}{c|}{209.83} & \multicolumn{1}{c|}{-11.84} & \multicolumn{1}{c|}{1.7} & \multicolumn{1}{c|}{bcu} \\ \midrule
\multicolumn{1}{|c|}{ID} & \multicolumn{1}{c|}{Energy (TeV)} & \multicolumn{1}{c|}{MJD} & \multicolumn{1}{c|}{RA (deg)} & \multicolumn{1}{c|}{Dec (deg)} & \multicolumn{1}{c|}{Ang Err (deg)} & \multicolumn{1}{c|}{} \\ \midrule
\multicolumn{1}{|c|}{38} & \multicolumn{1}{c|}{200.5} & \multicolumn{1}{c|}{56470.11} & \multicolumn{1}{c|}{93.3} & \multicolumn{1}{c|}{14} & \multicolumn{1}{c|}{1.2} & \multicolumn{1}{c|}{} \\ \midrule
\multicolumn{2}{|c|}{4FLG ASSOC} & \multicolumn{1}{c|}{ASSOC} & \multicolumn{1}{c|}{RA (deg)} & \multicolumn{1}{c|}{Dec (deg)} & \multicolumn{1}{c|}{Ang Sep (deg)} & \multicolumn{1}{c|}{Type} \\ \midrule
\multicolumn{2}{|c|}{*4FGL J0609.5+1402} & \multicolumn{1}{c|}{NVSS J060922+140744} & \multicolumn{1}{c|}{92.34} & \multicolumn{1}{c|}{14.12} & \multicolumn{1}{c|}{0.87} & \multicolumn{1}{c|}{bcu} \\ \midrule
\multicolumn{1}{|c|}{ID} & \multicolumn{1}{c|}{Energy (TeV)} & \multicolumn{1}{c|}{MJD} & \multicolumn{1}{c|}{RA (deg)} & \multicolumn{1}{c|}{Dec (deg)} & \multicolumn{1}{c|}{Ang Err (deg)} & \multicolumn{1}{c|}{} \\ \midrule
\multicolumn{1}{|c|}{44} & \multicolumn{1}{c|}{84.6} & \multicolumn{1}{c|}{56671.87} & \multicolumn{1}{c|}{336.7} & \multicolumn{1}{c|}{0} & \multicolumn{1}{c|}{1.2} & \multicolumn{1}{c|}{} \\ \midrule
\multicolumn{2}{|c|}{4FLG ASSOC} & \multicolumn{1}{c|}{ASSOC} & \multicolumn{1}{c|}{RA (deg)} & \multicolumn{1}{c|}{Dec (deg)} & \multicolumn{1}{c|}{Ang Sep (deg)} & \multicolumn{1}{c|}{Type} \\ \midrule
\multicolumn{2}{|c|}{4FGL J2227.9+0036} & \multicolumn{1}{c|}{PMN J2227+0037} & \multicolumn{1}{c|}{336.99} & \multicolumn{1}{c|}{0.61} & \multicolumn{1}{c|}{0.68} & \multicolumn{1}{c|}{bll} \\ \midrule
\multicolumn{1}{|c|}{ID} & \multicolumn{1}{c|}{Energy (TeV)} & \multicolumn{1}{c|}{MJD} & \multicolumn{1}{c|}{RA (deg)} & \multicolumn{1}{c|}{Dec (deg)} & \multicolumn{1}{c|}{Ang Err (deg)} & \multicolumn{1}{c|}{} \\ \midrule
\multicolumn{1}{|c|}{58} & \multicolumn{1}{c|}{52.6} & \multicolumn{1}{c|}{56859.75} & \multicolumn{1}{c|}{102.1} & \multicolumn{1}{c|}{-32.4} & \multicolumn{1}{c|}{1.3} & \multicolumn{1}{c|}{} \\ \midrule
\multicolumn{2}{|c|}{4FLG ASSOC} & \multicolumn{1}{c|}{ASSOC} & \multicolumn{1}{c|}{RA (deg)} & \multicolumn{1}{c|}{Dec (deg)} & \multicolumn{1}{c|}{Ang Sep (deg)} & \multicolumn{1}{c|}{Type} \\ \midrule
\multicolumn{2}{|c|}{4FGL J0649.5-3139} & \multicolumn{1}{c|}{NVSS J064933-313917} & \multicolumn{1}{c|}{102.39} & \multicolumn{1}{c|}{-31.65} & \multicolumn{1}{c|}{0.78} & \multicolumn{1}{c|}{bll} \\ \midrule
\multicolumn{1}{|c|}{ID} & \multicolumn{1}{c|}{Energy (TeV)} & \multicolumn{1}{c|}{MJD} & \multicolumn{1}{c|}{RA (deg)} & \multicolumn{1}{c|}{Dec (deg)} & \multicolumn{1}{c|}{Ang Err (deg)} & \multicolumn{1}{c|}{} \\ \midrule
\multicolumn{1}{|c|}{62} & \multicolumn{1}{c|}{75.8} & \multicolumn{1}{c|}{56987.77} & \multicolumn{1}{c|}{187.9} & \multicolumn{1}{c|}{13.3} & \multicolumn{1}{c|}{1.3} & \multicolumn{1}{c|}{} \\ \midrule
\multicolumn{2}{|c|}{4FLG ASSOC} & \multicolumn{1}{c|}{ASSOC} & \multicolumn{1}{c|}{RA (deg)} & \multicolumn{1}{c|}{Dec (deg)} & \multicolumn{1}{c|}{Ang Sep (deg)} & \multicolumn{1}{c|}{Type} \\ \midrule
\multicolumn{2}{|c|}{4FGL J1231.5+1421} & \multicolumn{1}{c|}{GB6 J1231+1421} & \multicolumn{1}{c|}{187.84} & \multicolumn{1}{c|}{14.35} & \multicolumn{1}{c|}{1.05} & \multicolumn{1}{c|}{bll} \\ \midrule
\multicolumn{1}{|c|}{ID} & \multicolumn{1}{c|}{Energy (TeV)} & \multicolumn{1}{c|}{MJD} & \multicolumn{1}{c|}{RA (deg)} & \multicolumn{1}{c|}{Dec (deg)} & \multicolumn{1}{c|}{Ang Err (deg)} & \multicolumn{1}{c|}{} \\ \midrule
\multicolumn{1}{|c|}{63} & \multicolumn{1}{c|}{97.4} & \multicolumn{1}{c|}{57000.14} & \multicolumn{1}{c|}{160} & \multicolumn{1}{c|}{6.5} & \multicolumn{1}{c|}{1.2} & \multicolumn{1}{c|}{} \\ \midrule
\multicolumn{2}{|c|}{4FLG ASSOC} & \multicolumn{1}{c|}{ASSOC} & \multicolumn{1}{c|}{RA (deg)} & \multicolumn{1}{c|}{Dec (deg)} & \multicolumn{1}{c|}{Ang Sep (deg)} & \multicolumn{1}{c|}{Type} \\ \midrule
\multicolumn{2}{|c|}{4FGL J1039.6+0535} & \multicolumn{1}{c|}{NVSS J103940+053608} & \multicolumn{1}{c|}{159.91} & \multicolumn{1}{c|}{5.6} & \multicolumn{1}{c|}{0.91} & \multicolumn{1}{c|}{bcu} \\ \midrule
\multicolumn{2}{|c|}{4FGL J1040.5+0617} & \multicolumn{1}{c|}{GB6 J1040+0617} & \multicolumn{1}{c|}{160.13} & \multicolumn{1}{c|}{6.28} & \multicolumn{1}{c|}{0.26} & \multicolumn{1}{c|}{bll} \\ \midrule
\multicolumn{2}{|c|}{4FGL J1043.6+0654} & \multicolumn{1}{c|}{NVSS J104323+065307} & \multicolumn{1}{c|}{160.84} & \multicolumn{1}{c|}{6.88} & \multicolumn{1}{c|}{0.98} & \multicolumn{1}{c|}{bll} \\ \midrule
\multicolumn{1}{|c|}{ID} & \multicolumn{1}{c|}{Energy (TeV)} & \multicolumn{1}{c|}{MJD} & \multicolumn{1}{c|}{RA (deg)} & \multicolumn{1}{c|}{Dec (deg)} & \multicolumn{1}{c|}{Ang Err (deg)} & \multicolumn{1}{c|}{} \\ \midrule
\multicolumn{1}{|c|}{71} & \multicolumn{1}{c|}{73.5} & \multicolumn{1}{c|}{57140.47} & \multicolumn{1}{c|}{80.7} & \multicolumn{1}{c|}{-20.8} & \multicolumn{1}{c|}{1.2} & \multicolumn{1}{c|}{} \\ \midrule
\multicolumn{2}{|c|}{4FLG ASSOC} & \multicolumn{1}{c|}{ASSOC} & \multicolumn{1}{c|}{RA (deg)} & \multicolumn{1}{c|}{Dec (deg)} & \multicolumn{1}{c|}{Ang Sep (deg)} & \multicolumn{1}{c|}{Type} \\ \midrule
\multicolumn{2}{|c|}{4FGL J0525.6-2008} & \multicolumn{1}{c|}{PMN J0525-2010} & \multicolumn{1}{c|}{81.36} & \multicolumn{1}{c|}{-20.18} & \multicolumn{1}{c|}{0.94} & \multicolumn{1}{c|}{bcu} \\ \midrule
\multicolumn{7}{r}{\multirow{2}{*}{EHE}} \\
\multicolumn{7}{r}{} \\ \midrule
\multicolumn{1}{|c|}{ID} & \multicolumn{1}{c|}{Energy (TeV)} & \multicolumn{1}{c|}{MJD} & \multicolumn{1}{c|}{RA (deg)} & \multicolumn{1}{c|}{Dec (deg)} & \multicolumn{1}{c|}{Ang Err (deg)} & \multicolumn{1}{c|}{} \\ \midrule
\multicolumn{1}{|c|}{13} & \multicolumn{1}{c|}{300} & \multicolumn{1}{c|}{55722.43} & \multicolumn{1}{c|}{272.22} & \multicolumn{1}{|c|}{35.55} & \multicolumn{1}{c|}{-} & \multicolumn{1}{c|}{} \\ \midrule
\multicolumn{2}{|c|}{4FLG ASSOC} & \multicolumn{1}{c|}{ASSOC} & \multicolumn{1}{c|}{RA (deg)} & \multicolumn{1}{c|}{Dec (deg)} & \multicolumn{1}{|c|}{Ang Sep (deg)} & \multicolumn{1}{c|}{Type} \\ \midrule
\multicolumn{2}{|c|}{4FGL J1808.8+3522} & \multicolumn{1}{c|}{2MASX   J18084968+3520426} & \multicolumn{1}{c|}{272.2} & \multicolumn{1}{c|}{35.34} & \multicolumn{1}{c|}{0.17} & \multicolumn{1}{c|}{bcu}  \\  \bottomrule
\end{tabular}
}
\caption{Correlations found between high-energy neutrinos reported in HESE and EHE catalogs and gamma-rays sources. Using the coordinates (RA and Dec) with the respective uncertainties,  we report blazars of the 4LAC associated a each neutrino. The sources marked with a * means that this source belongs to the Low Latitude sample.}
\label{tab:CorrHESEEHE}
\end{table}



\begin{table}[]
\centering
\scalebox{0.6}{
\begin{tabular}{@{}|c|c|c|c|c|c|c|c|@{}}
\toprule
\multicolumn{8}{r}{\multirow{2}{*}{Amon-HESE}} \\
\multicolumn{8}{r}{} \\ \midrule
ID & GCN NAME & Energy (TeV) & MJD & RA (deg) & Dec (deg) & Ang Err (deg) &  \\ \midrule
766165\_132518 & IceCube-190504A & - & 58607.77 & 65.78 & -37.44 & 1.23 &  \\ \midrule
\multicolumn{3}{|c|}{4FLG ASSOC} & ASSOC & RA (deg) & Dec (deg) & Ang Sep (deg) & Type \\ \midrule
\multicolumn{3}{|c|}{4FGL J0420.3-3745} & NVSS J042025-374443 & 65.1 & -37.74 & 0.63 & bcu \\ \midrule
\multicolumn{3}{|c|}{4FGL J0428.6-3756} & PKS 0426-380 & 67.17 & -37.94 & 1.2 & bll \\ \midrule
ID & GCN NAME & Energy (TeV) & MJD & RA (deg) & Dec (deg) & Ang Err (deg) &  \\ \midrule
66688965\_132229 & IceCube-190221A & - & 58535.35 & 267.36 & -16.93 & 1.23 &  \\ \midrule
\multicolumn{3}{|c|}{4FLG ASSOC} & ASSOC & RA (deg) & Dec (deg) & Ang Sep (deg) & Type \\ \midrule
\multicolumn{3}{|c|}{*4FGL J1744.9-1727} & 1RXS J174459.5-172640 & 266.23 & -17.45 & 1.2 & bcu \\ \midrule
\multicolumn{3}{|c|}{*4FGL J1751.6-1750} & NVSS J175120-175112 & 267.9 & -17.83 & 1.04 & bcu \\ \midrule
ID & GCN NAME & Energy (TeV) & MJD & RA (deg) & Dec (deg) & Ang Err (deg) &  \\ \midrule
12296708\_131624 & IceCube-181014A & - & 58405.49 & 225.18 & -34.79 & 1.23 &  \\ \midrule
\multicolumn{3}{|c|}{4FLG ASSOC} & ASSOC & RA (deg) & Dec (deg) & Ang Sep (deg) & Type \\ \midrule
\multicolumn{3}{|c|}{4FGL J1505.0-3433} & PMN J1505-3432 & 226.25 & -34.55 & 0.91 & bll \\ \midrule
ID & GCN NAME & Energy (TeV) & MJD & RA (deg) & Dec (deg) & Ang Err (deg) &  \\ \midrule
38561326\_128672 & IceCube-161103 & - & 57695.38 & 40.82 & 12.55 & 1.1 &  \\ \midrule
\multicolumn{3}{|c|}{4FLG ASSOC} & ASSOC & RA (deg) & Dec (deg) & Ang Sep (deg) & Type \\ \midrule
\multicolumn{3}{|c|}{4FGL J0244.7+1316} & GB6 J0244+1320 & 41.19 & 13.27 & 0.8 & bcu \\ \midrule
 &  &  &  &  &  &  &  \\ \midrule
\multicolumn{8}{ r }{\multirow{2}{*}{Amon-EHE}} \\
\multicolumn{8}{ r }{} \\ \midrule
ID & GCN NAME & Energy (TeV) & MJD & RA (deg) & Dec (deg) & Ang Err (deg) &  \\ \midrule
50579430\_130033 & IceCube-170922A & 119.98 & 58018.87 & 77.28 & 5.75 & 0.25 &  \\ \midrule
\multicolumn{3}{|c|}{4FLG ASSOC} & ASSOC & RA (deg) & Dec (deg) & Ang Sep (deg) & Type \\ \midrule
\multicolumn{3}{|c|}{4FGL J0509.4+0542} & TXS 0506+056 & 77.35 & 5.7 & 0.09 & bll \\ \bottomrule
\end{tabular}
}
\caption{Same as table 1, but with those neutrinos reported via AMON consortium.}
\label{tab:CorrAMON}
\end{table}


\begin{table}[]
\centering
\scalebox{0.6}{
\begin{tabular}{@{}|c|c|c|c|c|c|c|c|@{}}
\toprule
\multicolumn{8}{ r }{\multirow{2}{*}{Golden - Alerts}} \\
\multicolumn{8}{ r }{} \\ \midrule
ID & GCN NAME & Energy (TeV) & MJD & RA (deg) & Dec (deg) & Ang Err (deg) &  \\ \midrule
134698\_40735501 & IceCube-201114A & 214.29 & 59167.63 & 105.25 & 6.04 & 1.08 &  \\ \midrule
\multicolumn{3}{|c|}{4FLG ASSOC} & ASSOC & RA (deg) & Dec (deg) & Ang Sep (deg) & Type \\ \midrule
\multicolumn{3}{|c|}{*4FGL J0658.6+0636} & NVSS J065844+063711 & 104.64 & 6.6 & 0.81 & bcu \\ \midrule

\multicolumn{8}{ r }{\multirow{2}{*}{Bronze -   Alerts}} \\
\multicolumn{8}{ r }{} \\ \midrule
ID & GCN NAME & Energy (TeV) & MJD & RA (deg) & Dec (deg) & Ang Err (deg) &  \\ \midrule
134535\_41069485 & IceCube-200926B & 121.42 & 59118.94 & 184.75 & 32.929 & 1.81 &  \\ \midrule
\multicolumn{3}{|c|}{4FLG ASSOC} & ASSOC & RA (deg) & Dec (deg) & Ang Sep (deg) & Type \\ \midrule
\multicolumn{3}{|c|}{4FGL J1220.1+3432} & GB2 1217+348 & 185.04 & 34.53 & 1.63 & bll \\ \midrule
ID & GCN NAME & Energy (TeV) & MJD & RA (deg) & Dec (deg) & Ang Err (deg) &  \\ \midrule
133572\_82361476 & IceCube-191231A & 155.38 & 58849.46 & 46.36 & 20.42 & 1.72 &  \\ \midrule
\multicolumn{3}{|c|}{4FLG ASSOC} & ASSOC & RA (deg) & Dec (deg) & Ang Sep (deg) & Type \\ \midrule
\multicolumn{3}{|c|}{4FGL J0258.1+2030} & MG3 J025805+2029 & 44.54 & 20.51 & 1.71 & bll \\ \midrule
ID & GCN NAME & Energy (TeV) & MJD & RA (deg) & Dec (deg) & Ang Err (deg) &  \\ \midrule
133348\_80807014 & IceCube-191122A & 128.44 & 58809.95 & 27.25 & -0.04 & 1.82 &  \\ \midrule
\multicolumn{3}{|c|}{4FLG ASSOC} & ASSOC & RA (deg) & Dec (deg) & Ang Sep (deg) & Type \\ \midrule
\multicolumn{3}{|c|}{4FGL J0148.6+0127} & PMN J0148+0129 & 27.15 & 1.46 & 1.51 & bll \\ \bottomrule
\end{tabular}
}
\caption{Same as table 1, but with golden and bronze alerts.}
\label{tab:CorrGoldBronze}
\end{table}


\acknowledgments{
We acknowledge the support from Consejo Nacional de Ciencia y Tecnolog\'ia (CONACyT), M\'exico, grants IN106521. Antonio Galv\'an also wants to acknowledge the support of Becas Nacionales CONACyT.
}		

\bibliographystyle{JHEP} 
\bibliography{biblio}

\end{document}